\documentclass[letterpaper, 10pt, conference]{ieeeconf}

\IEEEoverridecommandlockouts                              
\usepackage{graphics} 
\usepackage{epsfig} 
\usepackage{times} 
\usepackage{amsmath} 
\usepackage{amssymb}  
\usepackage{cite}
\usepackage{bm}

\usepackage{color}
\usepackage{xcolor}

\usepackage{algorithm}
\usepackage{algorithmicx}
\usepackage{algpseudocode}
\algrenewcommand{\algorithmiccomment}[1]{\hskip3em\% #1}

\newtheorem{lemma}{Lemma}
\newtheorem{theorem}{Theorem}
\newtheorem{corollary}{Corollary}
\newtheorem{definition}{Definition}
\newtheorem{proposition}{Proposition}

\newcommand{\A}{\mathcal{A}}

\newcommand{\fC}{\mathfrak{C}}

\newcommand{\I}{\mathcal{I}}

\newcommand{\cP}{\mathcal{P}}

\newcommand{\R}{\mathbb{R}}

\newcommand{\rank}{\mathrm{rank\hspace{0.2ex}}}
\newcommand{\supp}[1]{{\rm supp}\left({#1}\right)}

\title{\LARGE \bf
	Data-Driven Resilience Assessment against Sparse Sensor Attacks
}

\author{Takumi Shinohara, Karl Henrik Johansson, and Henrik Sandberg
	\thanks{This work was supported in part by the Knut and Alice Wallenberg Foundation Wallenberg Scholar Grant, Swedish Research Council Distinguished Professor Grant (Project 2017-01078), Swedish Research Council (Project 2023-04770), Swedish Civil Defence and Resilience Agency (Project MAD-VAMCHS), and VINNOVA project ``Control-computing-communication co-design for autonomous industry (3C4AI)'' (Project 2025-01119).}
	\thanks{The authors are with the Department of Decision and Control Systems, KTH Royal Institute of Technology, and also with Digital Futures, 100 44 Stockholm, Sweden. (e-mail: tashin@kth.se, kallej@kth.se, hsan@kth.se).}}

\begin{document}
	
	\maketitle
	\thispagestyle{empty}
	\pagestyle{empty}
	
	\begin{abstract}
		We develop a data-driven framework for assessing the resilience of linear time-invariant systems against malicious false-data-injection sensor attacks.
		Leveraging sparse observability, we propose data-driven resilience metrics and derive necessary and sufficient conditions for two data-availability scenarios.
		For attack-free data, we show that when a rank condition holds, the resilience level can be computed exactly from the data alone, without prior knowledge of the system parameters.
		We then extend the analysis to the case where only poisoned data are available and show that the resulting assessment is necessarily conservative.
		For both scenarios, we provide algorithms for computing the proposed metrics and show that they can be computed in polynomial time under an additional spectral condition.
		A numerical example illustrates the efficacy and limitations of the proposed framework.
	\end{abstract}
	
	\section{Introduction}
	\label{section:introduction}
	Several recent studies have explored data-driven approaches to enhancing the resilience of control systems against attacks, demonstrating effective methods for detecting and/or identifying attacks.
	For instance, the papers \cite{2023TACShi,2025ECCTeixeira,2025TCS-ILi,2021TCYBLi} provide conditions and algorithms for detecting and identifying sensor attacks from attack-free data.
	The papers \cite{2020TII,2025AutomaticaChen} address the detection and identification problem against actuator attacks based on attack-free data.
	The data-driven attack detection problem against both sensor and actuator attacks is treated in \cite{2021L-CSSPasquialetti}.
	Furthermore, the paper \cite{2026TACLygeros} investigates the secure data reconstruction procedure under data manipulation, employing a behavioral approach.
	
	In this paper, we aim to quantitatively assess the level of resilience of control systems against sparse sensor attacks, using only state and output data.
	To the best of our knowledge, prior work on data-driven defense strategies has focused exclusively on attack detection and/or attack identification; how to assess the level of attack resilience directly from data has not been addressed.
	This gap matters because the resilience metric of control systems is essential for data-driven detection and identification against sparse sensor attacks, as suggested in \cite{2025ECCTeixeira}.
	Our primary contributions of this paper can be summarized as follows:
	\begin{enumerate}
		\item We propose data-driven resilience metrics against sensor attacks based on the notion of sparse observability.
		\item Using attack-free data, we derive a necessary and sufficient condition for assessing resilience and show that the exact resilience level can be computed from data when a rank condition holds.
		\item Using poisoned data, we derive a necessary and sufficient condition for assessing resilience and show that the resulting metric is necessarily conservative.
		\item For both scenarios, we analyze the computational complexity of computing the metrics and provide polynomial-time algorithms under a spectral condition.
	\end{enumerate}
	
	
	The rest of this paper is organized as follows.
	Section~\ref{section:problem} introduces the system and data models, the notions of model-based and data-driven sparse observability indices, and the problem formulation.
	Section~\ref{section:attack-free} considers the attack-free-data case, while Section~\ref{section:compromised} addresses the poisoned-data case.
	Section~\ref{section:simulation} presents a numerical example to illustrate the efficacy and limitations of the proposed framework, and Section~\ref{section:conclusion} concludes this paper.
	
	\subsubsection*{Notation}
	The notation $ |\I| $ is used to denote the cardinality of a set $ \I $.
	Denote by $ \mathfrak{C}^w_k $ the set of all $ k $-combinations from $ \{1, \ldots ,w\} $, i.e., all index sets in $ \{1, \ldots ,w\} $ whose cardinality is $ k $.
	For a vector, its support is defined as $ \supp{x} $.
	We use the notation $ \| x \|_0 \triangleq |\supp{x}\!|$ to denote the number of nonzero entries of a vector $ x $.
	We say a vector is $ \ell$-sparse if $ \| x \|_0 \leq \ell $.
	Also, for a matrix $ X \in \R^{m \times n} $, we say that $ X $ is $ \ell $-row-sparse if at most $ \ell $ rows are nonzero.
	Given a linear map $ A $, we use $ \ker A $ to denote the kernel of $ A $.
	The sets of eigenvalues and eigenvectors of a matrix $ A $ are, respectively, denoted by $ \sigma(A) $ and $ \mu (A) $.
	The Moore-Penrose pseudoinverse of any matrix $ A $ is denoted by $ A^\dagger $.
	The identity matrix of size $ n \times n $ is denoted by $ I_n $.
	Given two matrices $ A $ and $ B $ with the same number of columns, $ [A; B] $ denotes $ [A^\top, B^\top]^\top $.
	For a vector $ x \in \R^n $ and an index set $ \I \subseteq \{1,\ldots,n\} $, we use $ x_\I \in \R^{|\I|} $ to denote the subvector obtained from $ x $ by removing all elements except those indexed by $ \I $.
	Similarly, for a matrix $ A\in \R^{m \times n} $ and an index set $ \Gamma \subseteq \{1,\ldots,m\} $, we use $ A_\Gamma \in \R^{|\Gamma|\times n} $ to denote the submatrix obtained from $ A $ by removing all rows except those indexed by $ \Gamma $.
	
	\section{Problem Formulation}
	\label{section:problem}
	This section introduces the system and data models and reviews the notion of sparse observability, which underpins our resilience assessment. We then formalize the problem studied in this paper.
	
	\subsection{System Model}
	In this paper, we derive data-driven resilience metrics for the following linear time-invariant system:
	\begin{align}
		\label{eq:sytem_model}
		\left\lbrace
		\begin{array}{rl}
			x(k+1) \!\!\!&= \bar A x(k), \\
			y(k) \!\!\! &= \bar Cx(k) + a(k)= \widetilde y(k) + a(k),
		\end{array} \right.
	\end{align}
	where $ x(k) \in \R^n $ denotes the unknown system state, $ \widetilde y(k) \triangleq \bar Cx(k) \in \R^p  $ the nominal outputs, and $ y(k) \in \R^p $ the poisoned outputs.
	Assume that $ (\bar A,\bar C) $ is observable.
	Define $\mathcal{P} \triangleq \{1,\ldots,p\}$ as the index set of the outputs.
	
	The vector $ a(k) \in \R^p $ models an attack signal injected into the sensor outputs by an adversary.
	The attacker is assumed to be omniscient, meaning they possess complete knowledge of the system's state, nominal outputs, and system model.
	This paper also assumes that the attacker can generate an attack sequence $ a(k) $ arbitrarily with respect to stochastic properties, magnitude bounds, and time correlations based on their knowledge.
	Assume that the number of malicious signals is at most $ \ell $, i.e., $ \| a(k) \|_0 \leq \ell $, and that the set of compromised sensors is time-invariant, denoted by $ \Gamma^* \subseteq \mathcal{P} $ with $ |\Gamma^*| \leq \ell $.
	
	Since we are interested in data-driven analysis of system resilience, we assume that the system parameters $ \bar A $ and $ \bar C $ are unknown.
	We also assume that the attack vector $a(k)$ is unknown, whereas the state and output data\footnote{This paper considers the case where noiseless  state and output data are obtained, similar to \cite{2022CDCWaarde}.
		Analysis for the more general case where noisy input/output data are given is left for future work.} and the maximal attack number $ \ell $ are available.
	
	\subsection{Data Model}
	Following \cite{2020TACTrentelman}, the data model of (\ref{eq:sytem_model}) is given by
	\begin{align}
		\label{eq:X}
		X^+ & = \bar AX^-, \\
		\label{eq:Y}
		Y & = \bar CX^- + \bar E = \widetilde Y + \bar E,
	\end{align}
	with
	\begin{align*}
		X & \triangleq \left[\!\!
		\begin{array}{ccccc}
			x(0) & x(1) & \cdots & x(T-1) & x(T)
		\end{array}\!\!\right] \in \R^{n \times (T+1)}, \\
		X^+ & \triangleq \left[\!\!
		\begin{array}{ccccc}
			x(1) & x(2) & \cdots & x(T-1) & x(T)
		\end{array}\!\!\right] \in \R^{n \times T}, \\
		X^- & \triangleq \left[\!\!
		\begin{array}{ccccc}
			x(0) & x(1) & x(2) & \cdots & x(T-1)
		\end{array}\!\!\right] \in \R^{n \times T},\\
		Y & \triangleq \left[\!\!
		\begin{array}{ccccc}
			y(0) & y(1) & y(2) & \cdots & y(T-1)
		\end{array}\!\!\right] \in \R^{p \times T},\\
		\widetilde Y  & \triangleq \left[\!\!
		\begin{array}{ccccc}
			\widetilde y(0) & \widetilde y(1) & \widetilde y(2) & \cdots & \widetilde y(T-1)
		\end{array}\!\!\right] \in \R^{p \times T},\\
		\bar E & \triangleq \left[\!\!
		\begin{array}{ccccc}
			a(0) & a(1) & a(2) & \cdots & a(T-1)
		\end{array}\!\!\right] \in \R^{p \times T},
	\end{align*}
	where $ T \geq n $.
	Note that the attack matrix $ \bar E $ is $ \ell $-row-sparse.
	
	For the given data $ (X, Y) $, the set of systems consistent with the data can be obtained by
	\begin{align}
		\label{eq:sigma_attack}
		\Sigma & \triangleq \left\lbrace \!(A,C)\in\R^{n\times n} \!\times \!\R^{p \times n} \!:\! \left[\!\!
		\begin{array}{c}
			X^+ \\ Y
		\end{array}\!\!\right] \!= \!\left[\!\!
		\begin{array}{c}
			AX^- \\ CX^- + E
		\end{array}\!\!\right], \right. \nonumber \\
		&\left. \hspace{25mm}\exists E \in \R^{p\times T}, E\mathrm{~is~}\ell\text{-}\mathrm{row}\text{-}\mathrm{sparse}\right\rbrace.
	\end{align}
	Also, when the attack-free data $ (X, \widetilde Y) $ can be obtained, the set of systems is given by
	\begin{align}
		\label{eq:sigma_attack-free}
		\widetilde \Sigma & \triangleq \left\lbrace \!(A,C)\in\R^{n\times n} \!\times \!\R^{p \times n} \!:\! \left[\!\!
		\begin{array}{c}
			X^+ \\ \widetilde Y
		\end{array}\!\!\right] \!= \!\left[\!\!
		\begin{array}{c}
			A \\ C
		\end{array}\!\!\right]X^-\right\rbrace.
	\end{align}
	
	\subsection{Sparse Observability}
	For the system in the presence of sparse sensor attacks, the following notion of observability provides valuable insight into its ability to withstand adversarial intrusions \cite{2016TACTabuada,2022AutomaticaTabuada,2025OJ-CSYSShinohara}.
	\begin{definition}[Sparse Observability]
		The pair $ (\bar A,\bar C) $ is said to be \textit{$ \delta $-sparse observable} if $ (\bar A,\bar C_\Gamma) $ is observable for every set $ \Gamma \in \fC^{p}_{p-\delta} $, where $ \mathfrak{C}^p_{p-\delta} $ denotes the set of all $ (p-\delta) $-combinations from $ \{1, \ldots ,p\} $
		The largest integer $ \delta^{\max} $ such that the system is $ \delta^{\max} $-sparse observable is called the \textit{sparse observability index} for the system.
	\end{definition}
	
	A system is $ \delta $-sparse observable if it remains observable after removing any $ \delta $ sensors.
	In this paper, $ \delta^{\max} $ is unknown because the system model $ (\bar A,\bar C) $ is not given.
	
	Regarding a system in which at most $ \ell $ sensor attacks exist, the following properties have been established:
	\begin{itemize}
		\item The system state can be correctly estimated under $ \ell $-sparse sensor attacks if and only if the system is $ 2\ell $-sparse observable\footnote{The $ 2\ell $-sparse observability indicates that the system is $ \delta $-sparse observable with $ \delta = 2\ell $.} (see \cite[Proposition 2]{2014TACTabuada} or \cite[Theorem 3.2]{2016TACTabuada}).
		\item The sensor attack can be detected under $ \ell $-sparse sensor attacks if and only if the system is $ \ell $-sparse observable (see \cite[Theorem 16.1]{2020Tabuada}).
	\end{itemize}
	
	These established results imply that, to guarantee correct state estimation (or attack detection) in the presence of $ \ell $-sparse attacks, the system must remain observable even after any selection of $ 2\ell $ (or $ \ell $) sensors is removed.
	This means that the larger sparse observability index enables the state estimation and attack detection even in the presence of a large number of sensor attacks.
	Thus, the sparse observability index describes the system's resilience level against malicious sensor attacks \cite{2025OJ-CSYSShinohara}.
	Note that computing the sparse observability index of the system is coNP-hard \cite{2022AutomaticaTabuada}; therefore, no polynomial-time method is known even when the system model is given (unless $ \mathrm{P=coNP} $).
	
	For future analysis, we present different representations of the sparse observability.
	\begin{proposition}
		\label{proposition:sparse observability}
		For the system (\ref{eq:sytem_model}), the following conditions are equivalent:
		\begin{enumerate}
			\item The system $ (\bar A,\bar C) $ is $ \delta $-sparse observable.
			\item $ \rank[\bar A-\lambda I_n; \bar C_{\Gamma}] = n,~\forall \lambda \in \sigma(\bar A),~\Gamma \in \fC^{p}_{p-\delta} $.
			\item For any $ \Gamma \in \fC^{p}_{p-\delta} $, there is no nonzero vector $ v \in \mu(\bar A) $ such that $ \bar C_{\Gamma}v = 0 $.
			\item $ \| \bar Cv \|_0 > \delta,~\forall v \in \mu(\bar A) $.
		\end{enumerate}
	\end{proposition}
	\begin{proof}
		The equivalence follows from the observability rank test and the PBH test (see, e.g., \cite{Linear Systems}); hence, we omit the details.
	\end{proof}
	
	This paper addresses the problem of assessing the system's resilience against sparse sensor attacks using only state and output data.
	To this end, we first define the data informativity for sparse observability, referring to \cite{2022CDCWaarde}, as follows.
	\begin{definition}
		\label{definition:informative_sparse}
		We say that the data $ (X, Y) $ (resp. $ (X, \widetilde Y) $) are \textit{informative for $ \varrho $-sparse observability} if $ \Sigma \neq \emptyset $ (resp. $ \widetilde \Sigma \neq \emptyset $) and every pair $ (A,C) \in \Sigma $ (resp. $ (A,C) \in \widetilde \Sigma $) is $ \varrho $-sparse observable.
		The largest integer $ \varrho^{\max} $ for which the data are informative for $ \varrho^{\max} $-sparse observability is called the \textit{data-driven sparse observability index} for the system.
	\end{definition}
	
	\begin{figure}[t]
		\begin{center}
			\includegraphics[width=0.9\linewidth]{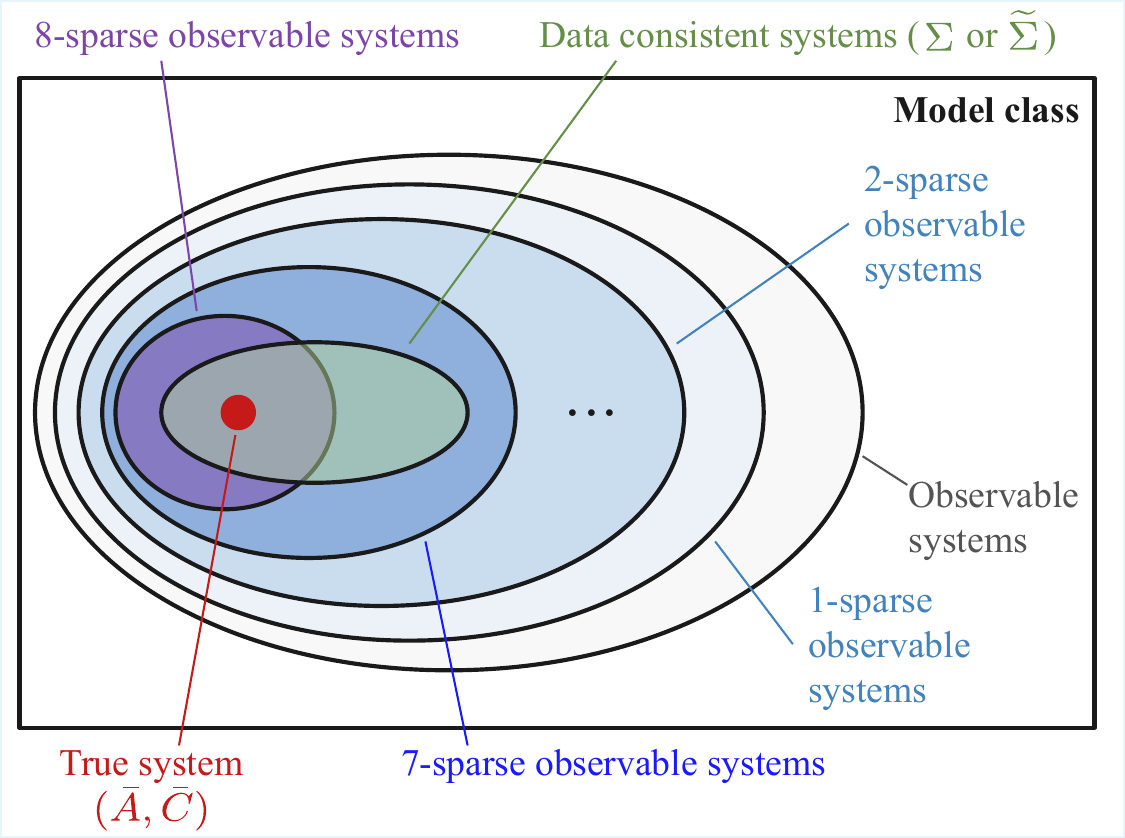}
			\vspace{-3.5mm}
			\caption{Illustration of data informativity for $ \delta $-sparse observability.
				In this illustration, the true but unknown system $ (\bar A,\bar C) $ is $ 8 $-sparse observable, i.e., $ \delta^{\max} = 8 $.
				On the other hand, the data are informative for $ 7 $-sparse observability, i.e., $ \varrho^{\max} = 7 $, because the data-consistent systems are $ 7 $-sparse observable. Thus, in this illustration, $ \varrho^{\max} < \delta^{\max} $.}
			\label{fig:sparse_observability}
		\end{center}
		\vspace{-6mm}
	\end{figure}
	
	This informativity notion indicates that all systems being consistent with the collected data (i.e., $ \Sigma $ or $ \widetilde{\Sigma} $) have the property of $ \varrho $-sparse observability.
	The data-driven sparse observability index refers to the largest value of the index that can be determined from the given data.
	As illustrated in Fig.~\ref{fig:sparse_observability}, regarding the relationship between $ \varrho^{\max} $ and $ \delta^{\max} $, we have $ \varrho^{\max} \leq \delta^{\max} $, which implies that there might be a gap between $ \delta^{\max} $ and $ \varrho^{\max} $.
	
	
	\subsection{Problem of Interest}
	Our goal is to provide data-driven conditions and algorithms to compute $ \varrho^{\max} $, thereby assessing the system's resilience to sparse sensor attacks.
	In the next section, we consider a scenario in which attack-free data $ (X,\widetilde Y) $ are available.
	Then, in Section \ref{section:compromised}, we address the case where only poisoned data $ (X, Y) $ are available.
	
	\section{Data-Driven Sparse Observability Index from Attack-free Data}
	\label{section:attack-free}
	In this section, we deal with the case where the attack-free data $ (X,\widetilde Y) $ are available.
	We first derive a necessary and sufficient condition for the data informativity for $ \varrho $-sparse observability.
	Using the condition, we then provide an algorithm to compute $\varrho^{\max}$ and discuss computational complexity.
	We also provide a polynomial-time algorithm under an additional spectral condition.
	
	\subsection{Necessary and Sufficient Condition}
	To derive the condition, we first show that the full row-rank property of $ X^- $ is a necessary condition for the data informativity for sparse observability.
	\begin{lemma}
		\label{lemma:X_full_row_rank}
		If the attack-free data $ (X,\widetilde Y) $ are informative for $ \varrho $-sparse observability for some nonnegative integer $ \varrho < p $, then $ X^- $ has full row rank, i.e., $ \rank X^- = n $.
	\end{lemma}
	\begin{proof}
		Due to space limitations, we omit the proof.
	\end{proof}
	
	The contrapositive of this lemma says if $  \rank X^- < n $, then the data $ (X,\widetilde Y) $ are not informative for $ \varrho $-sparse observability for any $ \varrho $.
	Using this lemma, we derive a necessary and sufficient condition for the data informativity for $ \varrho $-sparse observability.
	\begin{theorem}
		\label{theorem:informative_attack-free}
		The attack-free data $ (X,\widetilde Y) $ are informative for $ \varrho $-sparse observability if and only if
		\begin{align}
			\label{eq:informative_attack-free}
			\rank \left[\!\!
			\begin{array}{c}
				X^+ \!-\! \lambda X^- \\
				\widetilde  Y_{\Gamma}
			\end{array}\!\!\right] \!=\! n,~\forall \lambda \!\in \!\sigma \left(X^+ \left(X^-\right)^\dagger \right),~\forall \Gamma \!\in \!\fC^{p}_{p-\varrho}.
		\end{align}
	\end{theorem}
	\vspace{1.5mm}
	\begin{proof}
		Using \cite[Proposition 6]{2020TACTrentelman}, the system matrix $ \bar A $ is identifiable as $ \bar A = X^+ (X^-)^\dagger $.
		Hence, (\ref{eq:informative_attack-free}) can be written as
		\begin{align*}
			\rank \left[\!\!
			\begin{array}{c}
				\bar A\!-\! \lambda I_n \\
				C_{\Gamma}
			\end{array}\!\!\right]X^- \!=\! n,~\forall \lambda \!\in \!\sigma \left(\bar A \right),~\forall \Gamma\! \in \!\fC^{p}_{p-\varrho}
		\end{align*}
		for all $ (\bar A,C)\in \widetilde \Sigma $.
		Since $ \rank(AB)\leq \rank B $, we obtain $ n \leq \rank X^- $.
		Because $ X^- $ has $ n $ rows, this yields that $ \rank X^- = n $.
		Moreover, $ \rank(AB)\leq \rank A $ implies $ n \leq \rank [\bar A-\lambda I_n; C_{\Gamma}] $.
		Since this matrix has $ n $ columns, we conclude
		\begin{align*}
			n = \rank \left[
			\begin{array}{c}
				\bar A- \lambda I_n \\
				C_{\Gamma}
			\end{array}\right],~\forall \lambda \in \sigma(\bar A),~\forall \Gamma \in \fC^{p}_{p-\varrho},
		\end{align*}
		which implies, from Proposition~\ref{proposition:sparse observability}, that every pair $ (\bar A,C)\in \widetilde \Sigma $ is $ \varrho $-sparse observable.
		
		Next, for necessity, we argue by contradiction, and suppose that the data $ (X,\widetilde Y) $ are informative for $ \varrho $-sparse observability, but (\ref{eq:informative_attack-free}) does not hold.
		Then, there exist $ \lambda \in \sigma(\bar A)$ and $ \Gamma \in \fC^{p}_{p-\varrho} $ such that
		$ \rank [X^+ - \lambda X^-; \widetilde  Y_{\Gamma}] \neq n $.
		Equivalently, for these $ \lambda $ and $ \Gamma $, we have $ \rank([  \bar A- \lambda I_n;  C_{\Gamma}]X^-) \neq n $ for all $ (\bar A,C)\in\widetilde{\Sigma} $.
		Since $ \rank X^- = n $ from Lemma~\ref{lemma:X_full_row_rank}, this relation implies $ \rank [\bar A-\lambda I_n; C_{\Gamma}] < n $,
		so the system is not $ \varrho $-sparse observable.
		This contradicts the premise.
	\end{proof}
	
	This theorem provides a procedure for obtaining the data-driven sparse observability index $ \varrho^{\max} $.
	Specifically, this index can be computed by solving the following problem:
	%
	\begin{align}
		\label{eq:optimization_data-driven_clean_02}
		\varrho^{\max} =\max_{\varrho \in \{0,\ldots,p-1\}}~&\varrho~~~~\mathrm{s.t.}~\rank \!\!\left[\!\!\!
		\begin{array}{c}
			X^+ \!\!-\! \lambda X^- \\
			\widetilde  Y_{\Gamma}
		\end{array}\!\!\!\right] \!=\! n,\nonumber \\
		&\forall \lambda \!\in\! \sigma \left(\!X^+ \!\left(X^-\right)^\dagger \right)\!,~\forall \Gamma \!\in \!\fC^{p}_{p-\varrho}.
	\end{align}
	
	Algorithm~\ref{algorithm:attack-free} computes the data-driven sparse observability index.
	It is worth mentioning that this algorithm returns the exact sparse observability index $ \delta^{\max} $ if $ X^- $ has full row rank, as confirmed in the following corollary.
	
	\begin{corollary}
		\label{corollary:equivalence}
		Given the attack-free data $ (X,\widetilde Y) $, $ \delta^{\max} = \varrho^{\max} $ if and only if $ \rank X^- = n$.
	\end{corollary}
	\begin{proof}
		For sufficiency, if $ \rank X^- = n $, then
		\begin{align*}
			\rank \left[\!\!
			\begin{array}{c}
				X^+- \lambda X^- \\
				\widetilde{Y}_{\Gamma}
			\end{array}\!\!\right] = n~\Rightarrow~\rank \left[\!
			\begin{array}{c}
				\bar A- \lambda I_n \\
				C_{\Gamma}
			\end{array}\!\right] = n
		\end{align*}
		for all $ \lambda \in \sigma (\bar A )$, all $ \Gamma \in \fC^{p}_{p-\varrho} $, and all $ (\bar A,C)\in \widetilde{\Sigma} $.
		Hence, from Proposition~\ref{proposition:sparse observability}, the problem (\ref{eq:optimization_data-driven_clean_02}) yields the sparse observability index $ \delta^{\max} $.
		
		Conversely, for necessity, if $ \delta^{\max} = \varrho^{\max} $, then the attack-free data $ (X,\widetilde{Y}) $ are informative for $ \varrho $-sparse observability for some nonnegative integer $ \varrho $, which implies $\rank X^- = n$ by Lemma~\ref{lemma:X_full_row_rank}.
	\end{proof}
	
	Therefore, given attack-free data with full-row-rank $ X^- $, we can accurately compute the sparse observability index $ \delta^{\max} $ of the original system (\ref{eq:sytem_model}) by using Algorithm~\ref{algorithm:attack-free} and the attack-free data.
	This implies that the attack-resilience level of the system can be accurately assessed solely from the data, without relying on model parameters.
	
	\begin{algorithm}[t]
		\caption{Computation of data-driven sparse observability index from attack-free data $ (X,\widetilde{Y}) $}
		\label{algorithm:attack-free}
		\begin{algorithmic}[1]
			\Statex \hspace{-6mm} \textbf{Input:} $ X^- $ with $ \rank X^- =n $, $ X^+ $, $ \widetilde{Y}$, and $ p $
			\Statex \hspace{-6mm} \textbf{Output:} The data-driven sparse observability index $ \varrho^{\max} $
			\State Set $ \varrho = 1 $.
			\While{$ \varrho < p$}
			\For{all $ \Gamma \in \fC^{p}_{p-\varrho} $}
			\For{all $ \lambda \in \sigma (X^+ (X^-)^\dagger ) $}
			\State Compute $       \widetilde{r}_{(\Gamma,\lambda)} = \rank \left[\!\!
			\begin{array}{c}
				X^+ - \lambda X^- \\
				\widetilde  Y_{\Gamma}
			\end{array}\!\!\right] $.
			\EndFor
			\EndFor
			\If{$ \widetilde{r}_{(\Gamma,\lambda)} \!\!=\! n,~\!\forall \lambda\! \in\! \sigma (X^+ (X^-)^\dagger ) ,~\!\forall \Gamma\! \in \!\fC^{p}_{p-\varrho} $\!}
			\State Set $ \varrho = \varrho + 1 $.
			\Else
			\State \textbf{break};
			\EndIf
			\EndWhile
			\State \textbf{return:} $ \varrho^{\max} = \varrho -1$
		\end{algorithmic}
	\end{algorithm}
	
	\subsection{Complexity and Polynomial-Time Algorithm}
	From Corollary~\ref{corollary:equivalence}, if $ \rank X^- = n $, computing $ \varrho^{\max} $ is equivalent to computing $ \delta^{\max} $.
	As shown in \cite[Theorem 7]{2022AutomaticaTabuada}, the computation of $ \delta^{\max} $ is coNP-hard, and thus the computation of $ \varrho^{\max} $ via Algorithm~\ref{algorithm:attack-free} is also coNP-hard.
	This indicates that the problem is computationally intractable in general.
	However, under a specific condition, this computation can be executed in polynomial time, as described in the following proposition.
	\begin{proposition}
		\label{proposition:attack_free_polynomial}
		Given the attack-free data $ (X,\widetilde{Y}) $, if $ \rank X^- = n $ and every eigenvalue of $ X^+ \left(X^-\right)^\dagger $ has geometric multiplicity one,
		then $\varrho^{\max}$ can be computed in polynomial time.
	\end{proposition}
	\begin{proof}
		Due to space limitations, we omit the proof.
	\end{proof}
	
	Combining the results of Corollary~\ref{corollary:equivalence} and Proposition~\ref{proposition:attack_free_polynomial}, if $ X^- $ has full row rank and every eigenvalue of $ X^+ \left(X^-\right)^\dagger $ has geometric multiplicity one, then we can compute the exact sparse observability index only using the attack-free data in polynomial time.
	The polynomial-time algorithm is presented in Algorithm~\ref{algorithm:clean_polynomial}.
	Note that the tractability condition based on the geometric multiplicity of eigenvalues is consistent with the model-based result \cite{2022AutomaticaTabuada}: the general problem is intractable, whereas it can be computed in polynomial time when each eigenvalue has geometric multiplicity one.
	
	\begin{algorithm}[t]
		\caption{Polynomial-time computation of data-driven sparse observability index from attack-free data $ (X,\widetilde Y) $}
		\label{algorithm:clean_polynomial}
		\begin{algorithmic}[1]
			\Statex \hspace{-6mm} \textbf{Input:} $ X^- $ with $ \rank X^- =n $, $ X^+ $, $ \widetilde Y$, and $ p $
			\Statex \hspace{-6mm} \textbf{Output:} The data-driven sparse observability index $ \varrho^{\max} $
			\If{every eigenvalue of $ X^+ \left(X^-\right)^\dagger  $ has geometric multiplicity one}
			\For{all $ \lambda \in \sigma (X^+ (X^-)^\dagger ) $}
			\State Choose a unit eigenvector $ v $ such that
			\Statex $ \hspace{35mm}(X^+(X^-)^\dagger - \lambda I_n)v = 0 $.
			\State Compute $ z = \left(X^-\right)^\dagger v $ and $ \zeta_{(\lambda)} = \left\| \widetilde Yz\right\|_0 $.
			\EndFor
			\Else
			\State Use Algorithm~\ref{algorithm:attack-free}. \Comment{Not polynomial time.}
			\EndIf
			\State \textbf{return:} $ \varrho^{\max} =  \min_\lambda\{\zeta_{(\lambda)}\}-1  $
		\end{algorithmic}
	\end{algorithm}
	
	\section{Data-Driven Sparse Observability Index from Poisoned Data}
	\label{section:compromised}
	This section addresses the case where only poisoned data $ (X, Y) $ are available.
	In this setting, Theorem~\ref{theorem:informative_attack-free} is no longer applicable because the poisoning attacks can change the rank condition in (\ref{eq:informative_attack-free}).
	
	\subsection{Necessary and Sufficient Condition}
	We first derive a necessary and sufficient condition for the data informativity for $ \varrho $-sparse observability with the poisoned data $ (X,Y) $.
	To this end, define
	\begin{align}
		\A\triangleq \left\lbrace i \in \cP :Y_i \left(I- \Pi \right)\neq 0\right\rbrace,~~\Pi \triangleq \left(X^-\right)^\dagger X^-.
	\end{align}
	This set collects sensors that are certainly attacked given the observed data: since $ (CX^-)\left(I-\Pi\right)= 0 $, from (\ref{eq:Y}), it holds that $ Y(I-\Pi)=E(I-\Pi) $.
	Thus, if the $ i $th row satisfies $ Y_i (I-\Pi)\neq 0 $, then $ E_i (I-\Pi) \neq 0 $, namely, $E_i \neq 0$.
	This also implies, if $ |\A| > \ell $, then no $ \ell $-row-sparse explanation exists and hence $ \Sigma = \emptyset $.
	Using this set, we have the following result.
	\begin{theorem}
		\label{theorem:informative_compromised}
		Assume $ \rank X^- = n $ and $ |\A|\leq \ell $.
		Under $ \ell $-sparse sensor attacks, the poisoned data $ (X,Y) $ are informative for $ \varrho $-sparse observability if and only if
		\begin{align}
			\label{eq:informative_compromised}
			&\min_{\Gamma \supseteq \A, |\Gamma|\leq \ell}   \left\| Y_{\Gamma^\complement} z \right\|_0 > \varrho, \nonumber\\
			&\forall \lambda \in \sigma \left(X^+ \left(X^-\right)^\dagger \right),~\forall v \in \ker \left(\lambda I - X^+ \left(X^-\right)^\dagger \right)\setminus\{0\},
		\end{align}
		where $ \Gamma^\complement \triangleq \cP \setminus \Gamma $ and $ z \triangleq (X^-)^\dagger v $.
	\end{theorem}
	\begin{proof}
		Since $ \rank X^- = n $, the system matrix $ \bar A $ is identifiable as $ \bar A = X^+ \left(X^-\right)^\dagger $ \cite[Proposition 6]{2020TACTrentelman}.
		For necessity, fix any $ \Gamma \supseteq \A $ with $ |\Gamma|\leq \ell $.
		Define a matrix $ C \in \R^{p \times n} $ as follows:
		\begin{align}
			\label{eq:theorem_C}
			C_i \triangleq \left\lbrace
			\begin{array}{ll}
				Y_i\left(X^-\right)^\dagger, & i \in \Gamma^\complement, \\
				0, & i \in \Gamma.
			\end{array} \right.
		\end{align}
		Also, denote $ E \triangleq Y-CX^- $.
		Then, for $ i \in \Gamma^\complement $, because $ i \notin \A $, we have $ E_i = Y_i - C_i X^- = Y_i - Y_i \left(X^-\right)^\dagger X^- = Y_i \left(I-\Pi\right)=0 $,
		which yields
		\begin{align*}
			E_i = \left\lbrace
			\begin{array}{ll}
				0, & i \in \Gamma^\complement, \\
				Y_i - C_i X^- = Y_i, & i \in \Gamma.
			\end{array} \right.
		\end{align*}
		Since $ |\Gamma| \leq \ell $, the matrix $ E $ is $ \ell $-row-sparse, which implies $ (\bar A, C) \in \Sigma $.
		Hence, since the data $ (X,Y) $ are informative for $ \varrho $-sparse observability, the pair $ (\bar A, C) $ is $ \varrho $-sparse observable, which implies, from Proposition~\ref{proposition:sparse observability}, that $ \|C v\|_0 > \varrho,~\forall v \in \mu (\bar A)$.
		By (\ref{eq:theorem_C}), we obtain
		\begin{align*}
			\left\|C v\right\|_0  = \left\|C_{\Gamma^\complement} v\right\|_0 = \left\| Y_{\Gamma^\complement}\left(X^-\right)^\dagger v\right\|_0 = \left\| Y_{\Gamma^\complement}z\right\|_0 > \varrho.
		\end{align*}
		This relation holds for all $ \Gamma \supseteq \A $ with $ |\Gamma|\leq \ell $, and thus (\ref{eq:informative_compromised}) holds.
		
		For sufficiency, fix any $ (\bar A, C) \in \Sigma $.
		For this realization, denote the index set of the nonzero rows of $ E $ by $ \bar \Gamma \subseteq \cP $.
		Since $ E $ is $ \ell $-row-sparse, $ |\bar  \Gamma|\leq \ell $ and $ E_{\bar  \Gamma^\complement} = 0 $, where $ \bar \Gamma^\complement \triangleq \cP \setminus \bar \Gamma $.
		Also, we obtain $ Y_i (I-\Pi) = E_i - E_i (X^-)^\dagger X^- $ for all $ i \in \cP $, which implies $ Y_i(I-\Pi) = 0 $ for all $ i \in \bar \Gamma^\complement $.
		Thus, $ \bar \Gamma \supseteq \A $.
		From the relation $ Y = CX^- + E $, we obtain $ Y_{\bar \Gamma^\complement} = C_{\bar \Gamma^\complement}X^- $, and hence $ Y_{\bar\Gamma^\complement}z = C_{\bar\Gamma^\complement}v $.
		This yields that $ \|Cv\|_0 \geq \|C_{\bar\Gamma^\complement}v\|_0 = \|  Y_{\bar\Gamma^\complement}z \|_0 $.
		From (\ref{eq:informative_compromised}), we have $ \|  Y_{\Gamma^\complement}z \|_0 > \varrho $ for all $ \Gamma \supseteq \A $ with $|\Gamma|\leq \ell $, which implies $ \|Cv\|_0 > \varrho$.
		Thus, by Proposition~\ref{proposition:sparse observability}, $ (\bar A,C) $ is $ \varrho $-sparse observable, and therefore $ (X,Y) $ are data informative for $ \varrho $-sparse observable.
	\end{proof}

	This theorem shows that the poisoned data $ (X,Y) $ are informative for $ \varrho $-sparse observability when $ \|Y_{\Gamma^\complement}z\|_0 > \varrho $ (not $ \| Yz \|_0 > \varrho$) holds for all $ \Gamma \supseteq \A $ with $ |\Gamma|\leq \ell $.
	Considering that the model-based $ \delta $-sparse observability condition is characterized as $ \|\bar Cv\|_0 > \delta $ (c.f., Proposition~\ref{proposition:sparse observability}), the data-driven condition seems conservative.
	However, this data-driven condition is necessary to guarantee resilience under worst-case attacks.
	In a model-based setting, the measurement matrix $ C $ is given and hence the nominal measurement structure is known even in the presence of $ \ell $-sparse attacks.
	By contrast, in a model-free setting, we do not know the nominal measurement structure.
	That is, it is unclear whether each element of the poisoned data $ Y $ is a nominal or attacked output, and hence, even output data that have not actually been attacked must be interpreted as if they possibly have been attacked.
	Roughly speaking, in a model-free setting, sensor attacks not only affect the actually poisoned measurements but also cause \textit{misunderstandings} that make attack-free measurements appear as if they have been poisoned, which leads to the conservative condition.
	This highlights a limitation in assessing system resilience using poisoned data.
	An illustrative example in Section~\ref{section:simulation} provides further explanation.
	
	
	For (\ref{eq:informative_compromised}), since $ \Gamma \supseteq \A $, the vectors $ Y_{\Gamma^\complement} z $ only use indices in $ \A^\complement \triangleq \cP \setminus \A $.
	Hence, we may additionally remove up to $ \ell - |\A| $ indices from $ \A^\complement $ to minimize the number of nonzero entries, which yields that
	\begin{align*}
		\min_{\Gamma \supseteq \A, |\Gamma|\leq \ell}   \left\| Y_{\Gamma^\complement} z \right\|_0 = \max \left\lbrace0, \left\| Y_{\A^\complement} z \right\|_0 - (\ell-|\A|) \right\rbrace.
	\end{align*}
	Building on this transformation, the problem for calculating the data-driven sparse observability index $ \varrho^{\max} $ from the poisoned data can be described as
	\begin{align}
		\label{eq:optimization_data-driven_compromised}
		&\varrho^{\max} =\max_{\varrho \in \{0,\ldots,p-1\}}~~\varrho ~~~\mathrm{s.t.}~\lambda \in \sigma(\bar A):\nonumber \\
		&\min_{\substack{v \in \ker (\lambda I - \bar A)\\ v \neq 0}} \left\| Y_{\A^\complement}\left(X^-\right)^\dagger v\right\|_0 > \varrho + \left(\ell-|\A|\right),
	\end{align}
	where $ \bar A =  X^+ \left(X^-\right)^\dagger  $.
	Using this optimization problem, the algorithm for computing the data-driven sparse observability index from the poisoned data can be presented as Algorithm~\ref{algorithm:compromised}.
	
	\begin{algorithm}[t]
		\caption{Computation of data-driven sparse observability index from poisoned data $ (X,Y) $}
		\label{algorithm:compromised}
		\begin{algorithmic}[1]
			\Statex \hspace{-6mm} \textbf{Input:} $ X^- $ with $ \rank X^- =n $, $ X^+ $, $ Y$, $ p $, and $ \ell $
			\Statex \hspace{-6mm} \textbf{Output:} The data-driven sparse observability index $ \varrho^{\max} $
			\State Set $ \bar A = X^+ (X^-)^\dagger  $ and $ \Pi= (X^-)^\dagger X^- $.
			\State Set $ \A= \left\lbrace i \in \cP :Y_i \left(I- \Pi\right)\neq 0\right\rbrace $.
			\For{all $ \lambda \in \sigma (\bar A)$}
			\State Compute $ \zeta_{(\lambda)} = \min_{\substack{v \in \ker (\lambda I - \bar A)\\ v \neq 0}} \left\| Y_{\A^\complement}\left(X^-\right)^\dagger v\right\|_0$.
			\EndFor
			\State \textbf{return:} $ \varrho^{\max} =  \min_\lambda  \{\zeta_{(\lambda)}\} -(\ell - |\A|)- 1$
		\end{algorithmic}
	\end{algorithm}

	\subsection{Complexity and Polynomial-Time Algorithm}
	Since the inner problem in (\ref{eq:optimization_data-driven_compromised}) is an $ \ell_0 $ minimization, computing $ \varrho^{\max} $ via Algorithm~\ref{algorithm:compromised} is NP-hard in general.
	Under an additional spectral condition similar to Proposition~\ref{proposition:attack_free_polynomial}, if every eigenvalue of $ X^+ \left(X^-\right)^\dagger $ has geometric multiplicity one, then $\varrho^{\max}$ can be computed in polynomial time.
	\begin{proposition}
		\label{proposition:compromised_polynomial}
		Suppose that the poisoned data $ (X,Y) $ are given.
		If $ \rank X^- = n $ and every eigenvalue of $ X^+ \left(X^-\right)^\dagger $ has geometric multiplicity one, then $\varrho^{\max}$ can be computed in polynomial time.
	\end{proposition}
	\begin{proof}
		Due to space limitations, we omit the proof.
	\end{proof}
	
	This polynomial-time computation can be performed using Algorithm~\ref{algorithm:compromised_polynomial}.
	%
	%
	%
	%
	%
	
	\begin{algorithm}[t]
		\caption{Polynomial-time computation of data-driven sparse observability index from poisoned data $ (X,Y) $}
		\label{algorithm:compromised_polynomial}
		\begin{algorithmic}[1]
			\Statex \hspace{-6mm} \textbf{Input:} $ X^- $ with $ \rank X^- =n $, $ X^+ $, $ Y$, $ p $, and $ \ell $
			\Statex \hspace{-6mm} \textbf{Output:} The data-driven sparse observability index $ \varrho^{\max} $
			\State Set $ \bar A = X^+ (X^-)^\dagger  $ and $ \Pi = (X^-)^\dagger X^- $.
			\If{every eigenvalue of $ \bar A $ has geometric multiplicity one}
			\State Set $ \A= \left\lbrace i \in \cP :Y_i \left(I- \Pi \right)\neq 0\right\rbrace $.
			\For{all $ \lambda \in \sigma (\bar A)$}
			\State Choose a unit eigenvector $ v $ such that
			\Statex $ \hspace{11mm}(\bar A - \lambda I_n)v = 0 $.
			\State Compute $ z = \left(X^-\right)^\dagger v $.
			\State Compute $ \zeta_{(\lambda)} = \left\| Y_{\A^\complement}z\right\|_0 $.
			\EndFor
			\Else
			\State Use Algorithm~\ref{algorithm:compromised}. \Comment{Not polynomial time.}
			\EndIf
			\State \textbf{return:} $ \varrho^{\max} =  \min_\lambda  \{\zeta_{(\lambda)}\} -(\ell - |\A|)- 1$
		\end{algorithmic}
	\end{algorithm}

		\section{Numerical Example}
		\label{section:simulation}
		In this section, we illustrate the efficacy and limitations of the proposed framework using a pendulum system whose dynamics are given by
		\begin{align*}
			mL^2\ddot{\theta}+mLg\sin \theta = 0,
		\end{align*}
		where $ m $ is the mass of the pendulum, $ L $ its length, $ \theta $ the angle, and $ g $ the gravitational acceleration.
		In this example, we set $ m = 1 $, $ L = 1 $, and $ g = 9.8 $.
		For small angles of $ \theta $, $ \sin \theta \approx \theta $.
		In this case, the nonlinear second-order dynamics can be written as the first-order linear dynamics.
		For the dynamics, we assume the three sensors are deployed.
		Consequently, the continuous-time system is represented by
		\begin{align*}
			\dot{x} = \left[
			\begin{array}{cc}
				0 & 1 \\ -9.8 & 0
			\end{array}\right]x,~y = \left[
			\begin{array}{cc}
				1 & 0 \\
				1 & 1 \\
				0 & 1
			\end{array}\right] x,~x \triangleq \left[
			\begin{array}{cc}
				\theta & \dot{\theta}
			\end{array}\right]^\top.
		\end{align*}
		With a sampling time of $ 0.05~\mathrm{s} $, discretization yields the discrete-time model (\ref{eq:sytem_model}) with
		\begin{align*}
			\bar A = \left[
			\begin{array}{cc}
				0.9878 & 0.0498 \\ -0.4880 & 0.9878
			\end{array}\right],~\bar C=  \left[
			\begin{array}{cc}
				1 & 0 \\
				1 & 1 \\
				0 & 1
			\end{array}\right].
		\end{align*}
		The sparse observability index of this system can be computed as $ \delta^{\max} = 2 $, meaning that the system remains observable after removing any two sensors.
		
\begin{figure}[t]
			\begin{center}
				\includegraphics[width=\linewidth]{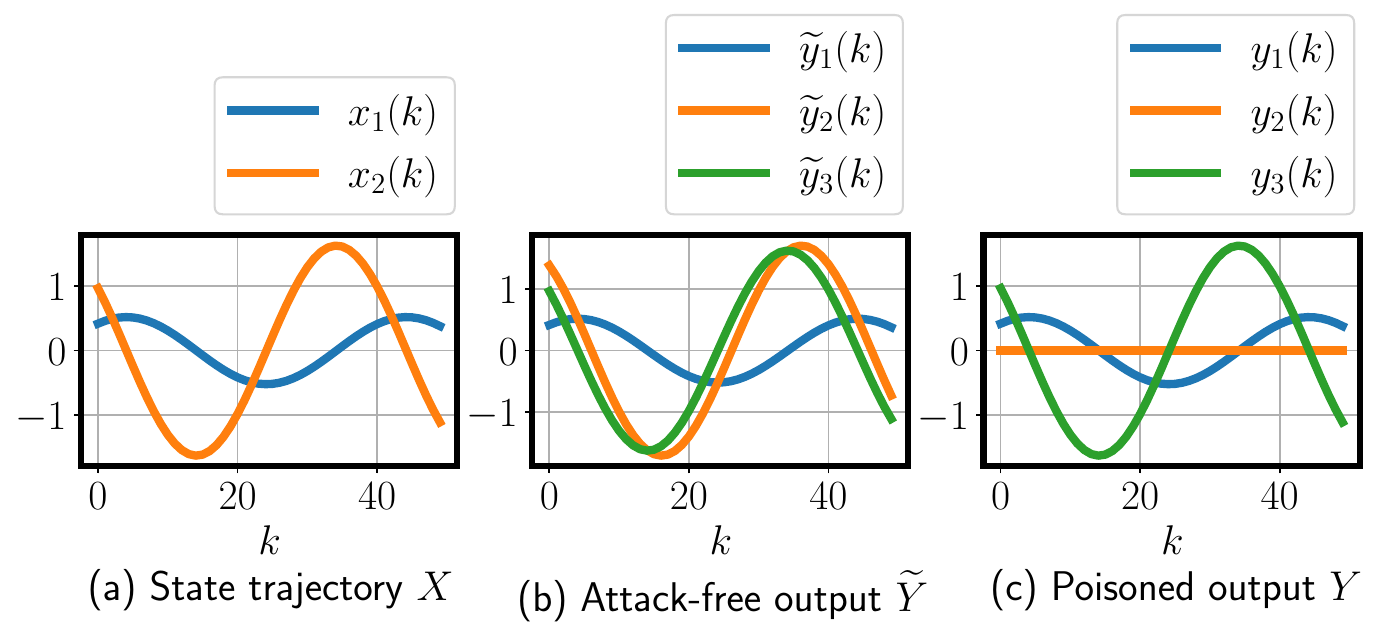}
				\vspace{-8.5mm}
				\caption{Simulation results for the pendulum system.}
				\label{fig:simulation_result}
			\end{center}
			\vspace{-6mm}
		\end{figure}

		We first consider the case where we have attack-free data $ (X, \widetilde{Y}) $, which are given in Figs.~\ref{fig:simulation_result}-(a) and \ref{fig:simulation_result}-(b), respectively.
		Then, we can obtain $ \varrho^{\max} = 2 $ via Algorithm~\ref{algorithm:attack-free}, which is equal to the exact sparse observability index $ \delta^{\max} $.
		Hence, the exact resilience level can be assessed using only the attack-free data.
		
		Next, consider the case of poisoned data $ (X, Y) $, which are shown in Figs.~\ref{fig:simulation_result}-(a) and \ref{fig:simulation_result}-(c), respectively, where a $ 1 $-sparse sensor attack is designed as $ a_1(k) = a_3(k) = 0 $ and $ a_2(k) = -\bar C_2x(k) $ so that $ y_2(k) = 0 $ for all $ k $.
		The number of attacked sensors is known ($ \ell = 1 $).
		Since only the poisoned data $ (X, Y) $ are available, the compromised sensor is unknown, and we must consider all data-consistent explanations.
		Assuming another measurement matrix $ C' \triangleq [0~0;0~0;0~1] $ and another $ 1 $-sparse attack sequence $ a_2(k) =a_3(k) = 0 $ and $ a_1(k) = \bar C_1x(k) $, these explain the same output $ Y $, which implies $ (\bar A,C')\in \Sigma $.
		This measurement matrix and attack sequence are not true, but there is no way to verify their correctness since only the data are now available.
		Therefore, under this spurious (yet data-consistent) explanation, the sparse observability index is zero, and thus the data-driven sparse observability index in this scenario is given by $ \varrho^{\max} = 0 $.
		This result can also be obtained from Algorithm~\ref{algorithm:compromised}.
		This example illustrates that, with poisoned data, resilience certification is conservative because attacks can distort the interpretation of even uncorrupted measurements.

		\section{Conclusion}
		\label{section:conclusion}
		We presented a data-driven framework based on sparse observability to assess the system's resilience against malicious sparse sensor attacks, using only the state and output data.
		For both attack-free and poisoned data, we derived the necessary and sufficient conditions for the data to be informative for $ \varrho $-sparse observability.
		Using these conditions, we developed algorithms to compute $ \varrho^{\max} $, which is a data-driven metric of attack resilience.
		Under an additional spectral condition, we provided polynomial-time procedures to compute $ \varrho^{\max} $.
		Finally, we illustrated the efficacy and limitations of our proposed framework through a numerical example.
		
		Although our results assume noiseless data, extending them to noisy data is a topic for future work.
		Another important direction is to develop analogous results based on the input/output data.

	\end{document}